\documentclass[a4paper,10pt]{scrartcl}

\usepackage{amsmath}
\usepackage{amsfonts}
\usepackage{amssymb}
\usepackage{graphicx}
\usepackage[english]{babel}
\usepackage{wasysym}
\usepackage{units}

\newcommand{\p}{\partial}
\newcommand{\reff}[1]{(\ref{#1})}
\newcommand{\vs}[1]{\vspace{#1mm}}
\newcommand{\vsO}{\vspace{.1cm}\hfill\\}
\newcommand{\vsT}{\vspace{.2cm}\hfill\\}

\title{\Large Lorentz Gauge Theory and Spinor Interaction}

\author{{\large N. Carlevaro}$^{\;a,\,b}$, {\large O.M. Lecian}$^{\;a,\,c}$ 
{\large and G. Montani}$^{\;a,\,c,\,d,\,e}$\vsT
\emph{\footnotesize $^a$ICRA -- International Center for Relativistic Astrophysics,}\vs{-2.5}\\
\emph{\footnotesize c/o Dep. of Physics - ``Sapienza'' Universit\`a di Roma}\\
\emph{\footnotesize $^b$Department of Physics, Polo Scientifico -- Universit\`a degli Studi di Firenze,}\vs{-2.5}\\
\emph{\footnotesize INFN -- Section of Florence, Via G. Sansone, 1 (50019), Sesto Fiorentino (FI), Italy}\\
\emph{\footnotesize $^c$ Department of Physics - ``Sapienza'' Universit\`a di Roma, Piazza A. Moro, 5 (00185), Rome, Italy}\\
\emph{\footnotesize $^d$ENEA -- C.R. Frascati (Department F.P.N.), Via Enrico Fermi, 45 (00044), Frascati (Rome), Italy}\\
\emph{\footnotesize $^{e}$ ICRANet -- C. C. Pescara, Piazzale della Repubblica, 10 (65100), Pescara, Italy}\\
{\footnotesize\ttfamily nakia.carlevaro@icra.it\quad lecian@icra.it\quad montani@icra.it}
}
\date{}
\begin{document}
\maketitle

%
\begin{abstract} \textbf{Abstract:} A gauge theory of the Lorentz group, based on the different behavior of spinors and vectors under local transformations, is formulated in a flat space-time and the role of the torsion field within the generalization to curved space-time is briefly discussed.

The spinor interaction with the new gauge field is then analyzed assuming the \emph{time gauge} and stationary solutions, in the non-relativistic limit, are treated to generalize the Pauli equation.

\vsO \emph{PACS}: 02.40.-k; 04.20.Fy; 04.50.+h; 11.15.-q
\end{abstract}

\section{Diffeomorphisms and Lorentz Transformations}
The gauge freedom of the gravitational interaction arises from the Principle of General Covariance and it is represented by the invariance under diffeomorphisms (\emph{diff.}'s). On the other hand, the equivalence of any local reference system deals with another gauge freedom expressed by local Lorentz transformations (\emph{Ltr.}'s). The main point is that the former gauge freedom reabsorbs the latter implying the nonexistence of independent connections of the local Lorentz group (LG). In fact, spin connections become only a particular combination of the tetrad fields and their derivatives and the local LG loses its status of independent gauge transformation
\begin{equation}
x^{\mu}\to x^{\mu}+\alpha^{\mu}_{\phantom1\nu}\left(
x\right)x^{\nu}=x^{\mu}+\tilde\alpha^{\mu}\left(x\right)\;.
\end{equation}
This scheme is well grounded in the case of scalar or macroscopic matter (spin-less matter) \cite{afldb}. In fact, from a mathematical point of view, using the tetrad formalism, \emph{diff.}'s act as a pullback on the tetrad, while local \emph{Ltr.}'s as a rotation of the local basis; in the case of isometric coordinate transformations, however, the latter can be restated in terms of the former. Under an isometric \emph{diff.}, spin connections transform like tensors, and cannot be gauge potentials for such \emph{diff.}-induced local rotations.

Conversely, if we deal with spin-$\nicefrac{1}{2}$ matter fields, local \emph{Ltr.}'s can no way be reabsorbed in the group of \emph{diff.}'s. This way, fermions give back to the LG its status of independent gauge group since they are described by a spinor representation of the LG, while the \emph{diff.} group does not admit any.
\vs{-5}

\section{Spinors and Lorentz Gauge Theory in Flat Space-time}
Using the tetrad formalism ($e^{a}_{\mu}$ being vierbein vectors), the Lagrangian density of spin-$\nicefrac{1}{2}$ fields on a \emph{4D} flat manifold reads
\begin{equation}\label{lagr}
\mathcal{L}=\tfrac{i}{2}\;\bar{\psi}\gamma^ae^{\mu}_a\p_{\mu}\psi-
\tfrac{i}{2}\;e^{\mu}_a\p_{\mu}\bar{\psi}\gamma^a\psi-m\,\bar{\psi}\psi\;,
\end{equation}
$\gamma_a$ being the Dirac matrices. Spinor fields have to recognize the isometric components of the \emph{diff.} as a local \emph{Ltr}. Therefore, let us now introduce a local \emph{Ltr.} $S=S(\Lambda(x))$: $\psi(x)\rightarrow S\psi(x)$ and assume that the $\gamma$ matrices transform locally like vectors ($S\,\gamma^{{a}}\,S^{-1}=(\Lambda^{-1})^{{a}}_{{b}}\,\gamma^{{b}}$). Taking into account the infinitesimal parameters $\epsilon_b^{a}(x)\ll1$, we define the usual relations for the LG
\begin{equation}
S=I-\tfrac{i}{4}\;\epsilon^{{a}{b}}\,\tau_{{a}{b}}\;,\qquad
\tau_{ab}=\tfrac{i}{2}\;[\gamma_{a},\gamma_{b}]\;,\qquad
[\tau_{cd},\tau_{ef}]=i\mathcal{F}^{\,ab}_{cdef}\,\tau_{ab}\;,
\end{equation}
$\tau_{ab}$ being group generators and $\mathcal{F}^{\,ab}_{cdef}$ the structure constants. In this scheme, the gauge invariance is restored by the covariant derivative
\begin{equation}
D_\mu\psi=(\p_\mu-\tfrac{i}{4}\,A_\mu)\,\psi=
(\p_\mu-\tfrac{i}{4}\,A_\mu^{ab}\,\tau_{ab})\,\psi\;,
\end{equation}
where the gauge transformation $\;\gamma^ae_{a}^{\mu}D_\mu\psi\to S\gamma^ae_{a}^{\mu}D_\mu\psi\;$ is assured by the gauge field $A_\mu=A^{ab}_\mu\tau_{ab}$ which transforms like a natural Yang-Mill field associated to the LG.

The considerations developed above can be generalized to curved space-time. In \cite{lecian}, a geometrical interpretation to the new Lorentz gauge field is provided, which can be identified with the tetradic projection of the \emph{contortion field} \cite{hehl}. The connections of the theory split up into two different terms: spin connections, which restore the Dirac algebra in the physical space-time and gauge connections, which guarantee the invariance under local \emph{Ltr.}'s, respectively. Comparing the results in flat space, gauge connections can be interpreted as real gauge fields of the local LG since they are non-vanishing quantities even in flat space-time, as requested for any gauge fields.

\section{Generalized Pauli Equation in Flat Space-time}
We now analyze the interaction of the \emph{4}-spinor $\psi$ with the new LG gauge field $A_\mu$. The implementation of the local Lorentz symmetry ($\partial_{\mu}\to D_{\mu}$) in flat spaces, leads to the Lagrangian density
\begin{equation}\label{lagrangian-tot}
\mathcal{L}=\tfrac{i}{2}\;\bar{\psi}\gamma^ae^{\mu}_a\p_{\mu}\psi-
\tfrac{i}{2}\;e^{\mu}_a\p_{\mu}\bar{\psi}\gamma^a\psi\,-\,m\,\bar{\psi}\psi\;+\;
\tfrac{1}{8}\,e^{\mu}_{c}\,\bar{\psi}\,\{\gamma^{c},
\tau_{ab}\}\,A^{ab}_{\mu}\,\psi\;,
\end{equation}
where $\{\gamma^c,\tau_{ab}\}=2\,\epsilon^{c}_{abd}\,\gamma_5\,\gamma^d$. To study the interaction term, let us now start from 
\begin{equation}
\mathcal{L}_{int}=\tfrac{1}{4}\;\bar{\psi}\;
\epsilon^{c}_{abd}\,\gamma_5\,\gamma^d\,A^{ab}_{c}\;\psi\;,
\end{equation}
where $a=\{0,i\}$, $i=\{1,2,3\}$. This allows us to split the gauge field into $A^{0i}_{0}\,,\;A^{ij}_{0}\,,\;A^{0i}_{\gamma}\,,
\;A^{ij}_{\gamma}$. We now impose the \emph{time-gauge} condition $A^{ij}_{0}\;=0$ and neglect the term $A^{0i}_{0}\;$ since it is summed over $\epsilon^{0}_{0i d}\equiv0$. By these considerations, the total Lagrangian density can be rewritten as 
\begin{equation}\label{lagrangian-split}
\mathcal{L}=\tfrac{i}{2}(\;\bar{\psi}\gamma^a\p_a\psi\,-\,
\p_a\bar{\psi}\gamma^a\psi)-m\,\bar{\psi}\psi
+\;\bar{\psi}\,C_0\,\gamma_5\gamma^0\psi\;+
\;\bar{\psi}\,C_i\,\gamma_5\gamma^i\psi\;,
\end{equation}
with the identifications
\begin{equation}
C_0=\tfrac{1}{4}\;\epsilon^{k}_{ij0}A^{i}_{k}\;,\qquad\quad
C_i=\tfrac{1}{4}\;\epsilon^{k}_{0ji}A^{0j}_{k}\;,
\end{equation}
where the component $C_0$ is related to rotations, while $C_i$ with boosts. Varying now the total action built up from the Lagrangian density \reff{lagrangian-split} with respect to $\psi^{\dagger}$, we get the following equation
\begin{equation}
(i\,\gamma^0\gamma^0\p_0\;+\;
C_i\,\gamma^0\gamma_5\gamma^i\;+\;
i\,\gamma^0\gamma^i\p_i\;+\;
C_0\,\gamma^0\gamma_5\gamma^0)\,\psi\;=\;m\,\gamma^0\,\psi\;,
\end{equation}
which is the eq. of motion for the \emph{4}-spinor $\psi$ interacting with the Lorentz-gauge field described here by the fields $C_0$ and $C_i$.

Let us now look for stationary solutions of the Dirac equation expressed by
\begin{equation}
\psi(\textbf{x},t)\to\psi(\textbf{x})\;e^{-i\mathcal{E}t}\;.
\end{equation}
The \emph{4}-component spinor $\psi(\textbf{x})$ can be expressed in terms of two \emph{2}-spinors $\chi(\textbf{x})$ and $\phi(\textbf{x})$ by writing\vs{-3}
\begin{equation}
\psi={\left(\begin{array}{c}\!\!\!\!\chi\!\!\!\!\\\phi\end{array}\right)}\;,\qquad\psi^\dagger=(\,\chi^\dagger\;,\;\phi^\dagger\,)\;,
\end{equation}
furthermore, let us assume the standard representation for the Dirac matrices
{\scriptsize\begin{equation}\nonumber
\gamma^i=\left(
\begin{array}{cc}
0 & \sigma_i \\
-\sigma_i & 0\\
\end{array}\right)\;,\qquad
\gamma^0=\left(
\begin{array}{cc}
\textbf{1} & 0 \\
0 & -\textbf{1}\\
\end{array}\right)\;,\qquad 
\gamma^5=\left(
\begin{array}{cc}
0 & \textbf{1}\\
\textbf{1} & 0\\
\end{array}\right)\;.
\end{equation}}Using such expressions, the \emph{2}-component spinors $\chi$ and $\phi$ are found to satisfy two coupled equations (here we write explicitly the \emph{3}-momentum $p^{\,i}$)
\begin{subequations}\label{stationary-eq}
\begin{align}
(\mathcal{E}-\sigma_i\cdot\,C^i)\,\chi\;
-\;(\sigma_i\cdot\,p^{\,i}+C_0)\,\phi\;&=\;m\,\chi\;,\label{stationary-eq1}\\
(\mathcal{E}-\sigma_i\cdot\,C^i)\,\phi\;
-\;(\sigma_i\cdot\,p^{\,i}+C_0)\,\chi\;&=-\;m\,\phi\;.\label{stationary-eq2}
\end{align}
\end{subequations}
These equations will be used below to study the non-relativistic limit of the Dirac-coupled equation \cite{shapiro}.

\paragraph{The Non-Relativistic Limit}\quad In order to investigate the low-energy limit, we write the spinor-field energy in the form
\begin{equation}
\mathcal{E}=E+m\;.
\end{equation}
Substituting this expression in the system \reff{stationary-eq}, the coupled equations rewrite now
\begin{subequations}
\begin{align}	
(E-\sigma_i\cdot\,C^i)\,\chi\;&=\;
(\sigma_i\cdot\,p^{\,i}+C_0)\,\phi\;,\label{non-rel-eq1}\\	(E-\sigma_i\cdot\,C^i+m)\,\phi\;&=\;
(\sigma_i\cdot\,p^{\,i}+C_0)\,\chi\;-\;m\,\phi\;.\label{non-rel-eq2}
\end{align}
\end{subequations}
In the non-relativistic limit, both $|E|$ and $|\,\sigma_i\cdot\,C^i|$ terms are small in comparison with the mass term $m$. Then, equation \reff{non-rel-eq2} can then be solved approximately as
\begin{equation}\label{small-components}
\phi\;=\;\tfrac{1}{2m}\;(\sigma_i\cdot\,p^{\,i}+C_0)\,\chi\;.
\end{equation}
It is immediate to see that $\phi$ is smaller than $\chi$ by a factor of order $\nicefrac{p}{m}$ (\emph{i.e.}, $\nicefrac{v}{c}$ where $v$ is the magnitude of the velocity). In this scheme, the \emph{2}-component spinors $\phi$ and $\chi$ form the so-called \emph{small} and \emph{large components}, respectively \cite{bransden}.

Substituting the expression of the small component \reff{small-components} in \reff{non-rel-eq1} and using standard Pauli matrices relations, we get 
\begin{equation}
E\,\chi\;=\;
\tfrac{1}{2m}\left[p^{2}\,+\,C_0^{2}\,+\,2\,C_0\,(\sigma_i\cdot\,p^{\,i})\,+\,
\sigma_i\cdot\,C^i\right]\,\chi\;.
\end{equation}
This equation exhibits a strong analogy with the electro-magnetic case and the so-called Pauli equation and can be used in the analysis of the energy levels as in the Zeeman effect. Let us now neglect the term $C_0^{2}$ and introduce a Coulomb potential $V(r)$, through the substitution $E\to E-V(r)$, obtaining the expressions
\begin{equation}
H_0=\frac{p^{2}}{2m}-\frac{Ze^{2}}{(4\pi \epsilon_0)r}\;,\qquad
H\,'=\tfrac{1}{2m}\left[\;2C_0\,(\sigma_i\cdot\,p^{\,i})\,+\,\sigma\cdot\,C^i\;\right]\;,
\end{equation}
which characterize the electron dynamics in a hydrogen-like atom in presence of a gauge field of the LG. It is worth noting the presence of a term related to the helicity of the \emph{2}-spinor: this coupling is controlled by the rotation-like component associated to $C_0$. A Zeeman-like coupling associated to the boost-like component $C_i$ is also present. The analysis of these interactions can be performed in a Stern-Gerlach thought experiment.


\end{document}